\documentclass[11pt]{article}

\usepackage{fullpage}
\usepackage{epsfig}
\usepackage{latexsym}
\usepackage{amsmath}
\usepackage{ifthen}
\usepackage{url}
\usepackage{indentfirst}

\newenvironment{proof}[1][0]{\noindent\ifthenelse{\equal{#1}{0}}{\textbf{Proof: }}{\textbf{Proof of #1: }}}{\hfill$\Box$}
\newtheorem{theorem}{Theorem}
\newtheorem{lemma}[theorem]{Lemma}

\newcommand{\work}{w}
\newcommand{\speed}{\sigma}

\newcommand{\after}{\text{after}}
\newcommand{\start}{S}
\newcommand{\next}{\text{succ}}

\newcommand{\proc}{\text{proc}}

\newcommand{\ALG}{\ensuremath{IncMerge}}

\begin{document}

\title{Power-aware scheduling for makespan and flow}

\author{David P. Bunde\thanks{Partially supported by NSF grant CCR 0093348.} \\
Department of Computer Science \\
Univ. Illinois at Urbana-Champaign \\
david.bunde@gmail.com}
\date{}
\maketitle

\begin{abstract}
We consider offline scheduling algorithms that incorporate speed scaling to
address the bicriteria problem of minimizing energy consumption and
a scheduling metric.
For makespan, we give linear-time algorithms to compute
all non-dominated solutions for the general uniprocessor
problem and for the multiprocessor problem when every job requires the
same amount of work.
We also show that the multiprocessor problem becomes NP-hard when jobs
can require different amounts of work.

For total flow, we show that the optimal flow corresponding to a
particular energy budget cannot be exactly computed on a machine
supporting arithmetic and the extraction of roots.
This hardness result holds even when scheduling equal-work jobs on a
uniprocessor.
We do, however, extend previous work by Pruhs et al. to give an
arbitrarily-good approximation for scheduling equal-work jobs on a
multiprocessor.
\end{abstract}

\section{Introduction}

Power consumption is becoming a major issue in computer systems.
This is most obvious for battery-powered systems such as laptops
because processor power consumption has been growing much more quickly
than battery capacity.
Even systems that do not rely on batteries have to deal with power
consumption since nearly all the energy consumed by a
processor is released as heat.
The heat generated by modern processors is becoming 
harder to dissipate and is particularly problematic
when large numbers of them are in close proximity, such as in
a supercomputer or a server farm.
The importance of the power problem has led to a great deal of
research on reducing processor power consumption; see overviews by
Mudge~\cite{mudge01}, Brooks et al.~\cite{brooks00},
and Tiwari et al.~\cite{tiwari98}.
We focus on the technique {\em dynamic voltage scaling}, which allows
the processor to enter low-voltage states.
Reducing the voltage reduces power consumption, but also forces a
reduction in clock frequency so the processor runs more slowly.
For this reason, dynamic voltage scaling is also called
{\em frequency scaling} and {\em speed scaling}.

This paper considers how to schedule processors with dynamic voltage
scaling so that the scheduling algorithm determines how fast to run
the processor in addition to choosing a job to run.
In classical scheduling problems, the input is 
a series of $n$ jobs $J_1, J_2, \ldots, J_n$.
Each job $J_i$ has a {\em release time} $r_i$, the earliest time it
can run, and a {\em processing time} $p_i$, the amount of
time it takes to complete.
With dynamic voltage scaling, the processing time is not known until
the schedule is constructed so instead each job $J_i$ comes with a
{\em work requirement} $\work_i$.
A processor running continuously at speed $\speed$ completes $\speed$
units of work per unit of time so job $J_i$ would have processing
time $\work_i/\speed$.
In general, a processor's speed is a function of time and the amount
of work it completes is the integral of this function over time.
This paper considers {\em offline} scheduling, meaning the algorithm
receives all the input together.
This is in constrast to {\em online} scheduling, where the algorithm
learns about each job at its release time.

To calculate the energy consumed by a schedule, we need a function
relating speed to power; the energy consumption is then the integral
of power over time.
Actual implementations of dynamic voltage scaling give a list of
speeds at which the processor can run.
For example, the AMD Athlon 64 can run at 2000MHz, 1800MHz, or
800MHz~\cite{athlon64power}.
Since the first work on power-aware scheduling
algorithms~\cite{weiser94b}, however, researchers
have assumed that the processor can run at an arbitrary
speed within some range.
The justification for allowing a continuous range of speeds is
twofold.
First, choosing the speed from a continuous range is an
approximation for a processor with a large number of possible speeds.
Second, a continuous range of possible clock speeds is observed by
individuals who use special motherboards to overclock their computers.

Most power-aware scheduling algorithms use the model
proposed by Yao et al.~\cite{yao95}, in which the processor can run at
any non-negative speed and $\text{power}=\text{speed}^\alpha$ for some
constant $\alpha>1$.
In this model, the energy required to run job $J_i$ at speed $\speed$
is $w_i\speed^{\alpha-1}$ since the running time is $w_i/\speed$.
This relationship between power and speed comes from an approximation
of a system's switching loss, the energy consumed by logic gates
switching values.

Most of our results do not assume a specific 
relationship between power and speed.
Except where otherwise stated, we just assume that power is a continuous,
strictly-convex function of processor speed.
Formally, strict convexity means that the line segment between any two
points on the power/speed curve lies above the curve except at its
endpoints.
More intuitively, strict convexity means that power increases
super-linearly with speed.
The power function is strictly convex when $\alpha > 1$ if
$\text{power}=\text{speed}^\alpha$.

To measure schedule quality, we use two classic metrics.
Let $S_i^A$ and $C_i^A$ denote the start and completion times of job
$J_i$ in schedule $A$.
Most of the paper focuses on minimizing the schedule's
{\em makespan}, $\max_i C_i^A$, the
completion time of the last job.
We also consider {\em total flow}, the sum over all jobs of
$C_i^A-r_i$, the time between the release and completion times of job
$J_i$. 

Either of these metrics can be improved by using more energy to speed
up the last job so the goals of low energy consumption and
high schedule quality are in opposition.
Thus, power-aware scheduling is a bicriteria optimization problem and
our goal becomes finding {\em non-dominated schedules}, those such
that no schedule can be both better and use less energy.
A common approach to bicriteria problems is to fix one of the
parameters.
In power-aware scheduling, this gives two interesting special cases.
If we fix energy, we get the
{\em laptop problem}, which asks ``What is the best schedule achievable
using a particular energy budget?''.
Fixing schedule quality gives the {\em server problem},
which asks ``What is the least energy required to achieve a desired
level of performance?''.

This paper considers both uniprocessor and multiprocessor scheduling.
In the multiprocessor setting, we assume that the processors have a
shared energy supply.
This corresponds to scheduling a laptop with a multi-core processor
or a server farm concerned only about total energy consumption and not
the consumption of each machine separately.

\paragraph*{Results}
Our results in power-aware scheduling are the following:
\begin{itemize}

\item
For uniprocessor makespan, we give an algorithm to find all 
non-dominated schedules. 
Its running time is linear once the jobs are sorted by arrival time.

\item
We show that there is no exact algorithm for uniprocessor total 
flow using arithmetic operations and the extraction of
$k^\text{th}$ roots.
This holds even with equal-work jobs.

\item
For a large class of ``reasonable'' scheduling metrics, we show how to extend 
uniprocessor algorithms to the multiprocessor setting with
equal-work jobs. 
Using this technique, we give an exact algorithm for multiprocessor
makespan of equal-work jobs and an
arbitrarily-good approximation
for multiprocessor total flow of equal-work jobs.

\item
We prove that multiprocessor makespan is NP-hard if jobs require
different amounts of work, even when all jobs arrive immediately.

\end{itemize}

The rest of the paper is organized as follows.
Section~\ref{background-section} describes related work.
Section~\ref{makespan-uni-section} gives the uniprocessor algorithm
for makespan.
Section~\ref{impossibility-section} shows that total flow cannot be
exactly minimized.
Section~\ref{multi-section} extends the uniprocessor results to
give multiprocessor algorithms for equal-work jobs and 
shows that general multiprocessor makespan is NP-hard.
Finally, Section~\ref{conc-section} discusses future work.

\section{Related work} \label{background-section}

The work most closely related to ours is due to
Uysal-Biyikoglu, Prabhakar, and El Gamal~\cite{uysalbiyikoglu02}, who
consider the problem of minimizing the energy of wireless 
transmissions.
This application has a totally different power function from those
occurring in dynamic voltage scaling, but their
algorithms only rely on the power function being continuous and
strictly convex.
They give a quadratic-time algorithm for the server version of
makespan.
Thus, our algorithm runs faster and also
finds all non-dominated schedules rather than just solving the
server problem.

El Gamal et al.~\cite{elgamal02} consider the wireless transmission
problem when the packets have different power functions, giving an
iterative algorithm that converges to an optimal solution.
They also show how to extend their algorithm to handle the case
when the buffer used to store active packets has bounded size and the
case when packets have individual deadlines.
Their algorithm can also be extended to schedule multiple
transmitters, but this does not correspond to a processor scheduling
problem.

Pruhs, van Stee, and Uthaisombut~\cite{pruhs05}
consider the laptop problem version of minimizing makespan for
jobs having precedence constraints where all jobs are released
immediately and $\text{power}=\text{speed}^\alpha$.
Their main observation, which they call the {\em power equality}, is
that the sum of the powers of the machines is
constant over time in the optimal schedule.
They use binary search to determine this value and then
reduce the problem to scheduling on related fixed-speed machines.
Previously-known~\cite{chudak97,chekuri01c} approximations for the
related fixed-speed machine problem then give an
$O(\log^{1+2/\alpha} m)$-approximation for power-aware makespan.
This technique cannot be applied in our setting because the power
equality does not hold for jobs with release dates.

Minimizing the makespan of tasks with precedence constraints has also been
studied in the context of project management.
Speed scaling is possible when additional resources
can be used to shorten some of the tasks.
Pinedo~\cite{pinedo05} gives heuristics for some
variations of this problem.

The only previous power-aware algorithm to minimize total flow is by
Pruhs, Uthaisombut, and Woeginger~\cite{pruhs04b},
who consider scheduling equal-work jobs on a uniprocessor.
In this setting, they observe that jobs can be run in order of 
release time and then prove the following relationships between the
speed of each job in the optimal solution:

\begin{theorem}[\cite{pruhs04b}]
\label{flow-relationship-theorem}
Let $J_1, J_2, \ldots, J_n$ be equal-work jobs ordered by release
time.
In the schedule $OPT$ minimizing total flow for a given energy budget
where $\text{power}=\text{speed}^\alpha$,
the speed $\speed_i$ of job $J_i$ (for $i \neq n$) obeys the following:
\begin{itemize}

\item
If $C_i^{OPT} < r_{i+1}$, then $\speed_i = \speed_n$.

\item
If $C_i^{OPT} > r_{i+1}$, then
$\speed_i^\alpha = \speed_{i+1}^\alpha + \speed_n^\alpha$.

\item
If $C_i^{OPT} = r_{i+1}$, then
$\speed_n^\alpha \leq \speed_i^\alpha \leq \speed_{i+1}^\alpha + \speed_n^\alpha$.
\end{itemize}
\end{theorem}

These relationships, together with observations about when
the optimal schedule changes configuration, give an algorithm
based on binary search that 
finds an arbitrarily-good approximation for either the laptop or the
server problem.
In fact, they can plot the exact tradeoff between total flow and
energy consumption for optimal schedules in which the third
relationship of Theorem~\ref{flow-relationship-theorem} does not occur.
Our impossibility result in Section~\ref{impossibility-section}
shows that the difficulty caused by the third relationship cannot be
avoided. 

The idea of power-aware scheduling was proposed by 
Weiser et al.~\cite{weiser94b}, who use trace-based simulations to
estimate how much energy could be saved by slowing the
processor to remove idle time.
Yao et al.~\cite{yao95} formalize this problem by assuming each job
has a deadline and seeking the minimum-energy schedule that satisfies
all deadlines. 
They give an optimal offline algorithm and propose two online
algorithms.
They show one is $(2^{\alpha-1}\alpha^\alpha)$-competitive, i.e. it
uses at most $2^{\alpha-1}\alpha^\alpha$ times the optimal energy.
Bansal et al.~\cite{bansal04c} analyze the other,
showing it is $\alpha^\alpha$-competitive.
Bansal et al.~\cite{bansal04c} also give another algorithm
that is $(2(\alpha/(\alpha-1))^\alpha e^\alpha)$-competitive.

Power-aware scheduling of jobs with deadlines has also been considered
with the goal of minimizing the CPU's maximum
temperature.
Bansal et al.~\cite{bansal04c} propose this problem and
give an offline solution based on convex programming.
Bansal and Pruhs~\cite{bansal05} analyze the online algorithms
discussed above in the context of minimizing maximum temperature.

A different variation is to assume that the processor can only choose
between discrete speeds.
Chen et al.~\cite{chen05} show that minimizing energy consumption
in this setting while meeting all deadlines is NP-hard,
but give approximations for some special cases. 

Another algorithmic approach to power management is to identify
times when the processor or parts of it can be partially or completely
powered down.
Irani and Pruhs~\cite{irani05} survey work along these lines as well
as approaches based on speed scaling.

\section{Makespan scheduling for a single processor}
\label{makespan-uni-section}

Our first result is an algorithm to find
all non-dominated schedules for uniprocessor power-aware makespan.
We begin by solving the laptop problem for an energy budget $E$.
Let $OPT$ be an optimal schedule for this problem, i.e. 
$OPT$ has minimum makespan among schedules using energy $E$.

\subsection{Algorithm for laptop problem}

To find $OPT$, we establish properties it must satisfy.
(We omit formal proofs for most of the properties and merely describe
the relevant ideas.)
Our first property is due to Yao, Demers, and Shenker~\cite{yao95},
who observed that the speed does not change during a job or energy
could be saved by running that job at its average speed.

\begin{lemma}[\cite{yao95}] \label{job-const-speed-lemma} \label{first-property-lemma}
Each job runs at a single speed in $OPT$.
\end{lemma}

We use $\speed_i^A$ to denote the speed of job $J_i$ in schedule $A$,
omitting the schedule when it is clear from context.

The second property allows us to fix the order in which jobs are run.

\begin{lemma} \label{fifo-lemma}
Without loss of generality, $OPT$ runs jobs in order of their
release times.
\end{lemma}

Lemma~\ref{fifo-lemma} holds because reordering jobs (without changing
their speeds) so that a job runs before jobs released after it
produces a legal schedule.
To simplify notation, we assume the jobs are indexed so
$r_1 \leq r_2 \leq r_3 \leq \ldots \leq r_n$.

The third property is that $OPT$ is not idle between the release of
the first job and the completion of the last job.

\begin{lemma} \label{no-idle-lemma}
$OPT$ is not idle between the release of job $J_1$ and the completion
of job $J_n$
\end{lemma}

Lemma~\ref{no-idle-lemma} holds because slowing down the job running
before a period of idle time saves energy, which can then be used to speed up
the last job and reduce the makespan.

Stating the next property requires a definition.
A {\em block} is a maximal substring of jobs
such that each job except the last finishes after the arrival of its
successor.
For brevity, we denote a block with the indices of its first and last
jobs.
Thus, the block with jobs $J_i, J_{i+1}, \ldots, J_{j-1}, J_j$ is
block $(i,j)$.
The fourth property is the analog of Lemma~\ref{job-const-speed-lemma}
for blocks.

\begin{lemma} \label{block-const-speed-lemma}
In $OPT$, jobs in the same block run at the same speed.
\end{lemma}

\begin{proof}
If the lemma does not hold, we can find two adjacent jobs $J_i$ and
$J_{i+1}$ in the same block of $OPT$ with $\speed_i \ne \speed_{i+1}$.
Let $\epsilon$ be a positive number less than the amount of work
remaining in job $J_i$ at time $r_{i+1}$.
Consider changing the schedule by running $\epsilon$ work of $J_i$ at
speed $\speed_{i+1}$ and $\epsilon$ work of $J_{i+1}$ at speed $\speed_i$.
Since the block contains the same amount of work at each speed, the
makespan is unchanged and the same amount of energy is used.
By construction, this change does not cause the schedule to violate
any release times.
Job $J_i$ does not run at a constant speed, however,
contradicting Lemma~\ref{job-const-speed-lemma}.
\end{proof}
\vspace*{1em}

Lemma~\ref{block-const-speed-lemma} shows that speed is a
property of blocks.
In fact, if we know how $OPT$ is broken into blocks, we can compute
the speed of each block.
The definition of a block and Lemma~\ref{no-idle-lemma} mean that
block $(i,j)$ starts at time $r_i$.
Similarly, block $(i,j)$ completes at time $r_{j+1}$ unless it is the
last block.
Thus, any block $(i,j)$ other than the last runs at speed
$(\sum_{k=i}^j \work_k)/(r_{j+1}-r_i)$.
To compute the speed of the last block, we subtract the energy used by
all the other blocks from the energy budget $E$.
We choose the speed of the last block to exactly use the
remaining energy.

Using the first four properties, an $O(n^2)$-time
dynamic programming algorithm can find the best way to divide
the jobs into blocks.
To improve on this, we establish the following restriction on
allowable block speeds:

\begin{lemma} \label{optimal-blocks-increasing-lemma} \label{last-property-lemma}
The block speeds in $OPT$ are non-decreasing.
\end{lemma}

\begin{proof}
Suppose to the contrary that $OPT$ runs a block $(i,j)$ faster
than block $(j+1,k)$.
Let $\epsilon>0$ be less than the amount of work in either block.
We modify the schedule by running $\epsilon$ of the work in each
block at the other block's speed.
This does not change when the pair of blocks complete or how much
energy they consume since the same amount of work is run at each
speed.
The modified schedule is valid since no job starts earlier than in $OPT$.
Thus, we have created another optimal schedule, but it runs block
$(i,j)$ at two speeds, contradicting either Lemma~\ref{job-const-speed-lemma}
or Lemma~\ref{block-const-speed-lemma}.
\end{proof}
\vspace*{1em}

It turns out that $OPT$ is the only schedule having all the
properties
given by Lemmas~\ref{first-property-lemma}--\ref{last-property-lemma}.

\begin{lemma} \label{blocks-increasing-uniqueness-lemma}
For any energy budget, there is a unique schedule
having the following properties:
\begin{enumerate}
\item Each job runs at a single speed
\item Jobs are run in order of release time
\item It is not idle between the release of job $J_1$ and the
  completion of job $J_n$
\item Jobs in the same block run at the same speed
\item The blocks speeds are non-decreasing
\end{enumerate}
\end{lemma}

\begin{proof}
Suppose to the contrary that $A$ and $B$ are different schedules
obeying all five properties and consuming the same amount of energy.
Since each schedule is determined by its blocks, $A$ and $B$ 
must have different blocks.
Without loss of generality, suppose the first difference occurs when
job $J_i$ is the last job in its block for schedule $A$ but not for
schedule $B$.
We claim that every job indexed at least $i$ runs slower
in schedule $B$ than in schedule $A$.
Since energy consumption increases with speed, this implies that
schedule $B$ uses less energy than schedule $A$, a contradiction.

In fact, we prove the strengthened claim that
every job indexed at least $i$ runs slower and finishes later
in schedule $B$ than in schedule $A$.
First, we show this holds for job $J_i$.
Job $J_i$ ends its block in schedule $A$ but not in
schedule $B$ so $C_i^B > r_{i+1} = C_i^A$.
Since each schedule begins the block containing job $J_i$ at the same
time and runs the same jobs before job $J_i$, 
job $J_i$ runs slower in schedule $B$ than schedule $A$.

Now we assume that the strengthened claim holds for jobs indexed below
$j$ and consider job $J_j$.
Since each job $J_i,\ldots,J_{j-1}$ finishes no earlier than its
successor's release time in schedule $A$, each finishes after 
its successor's release time in schedule $B$.
Thus, none of these jobs ends a block in schedule $B$ and schedule $B$
places jobs $J_i$ and $J_j$ in the same block,
which implies $\speed_j^B=\speed_i^B$.
Speed is non-decreasing in schedule $A$ so
$\speed_i^A \leq \speed_j^A$.
Therefore, $\speed_j^B = \speed_i^B < \speed_i^A \leq \speed_j^A$ so
job $J_j$ runs slower in schedule $B$ than in schedule $A$.
Job $J_j$ also finishes later because
job $J_{j-1}$ finishing later implies that job $J_j$ starts later.
\end{proof}

Because only $OPT$ has all five properties, we can solve
the laptop problem by finding a schedule with the properties.
For this task, we propose an algorithm \ALG.
This algorithm maintains a tentative list of blocks, initially empty.
Each block knows its speed, calculated as described above from the
release time of the next job (including jobs not yet
added to the schedule) or the energy budget.
Jobs are added to the schedule one at a time in order of their
release times.
When a new job is added, it starts in its own block.
Then, while the last block runs slower than its predecessor,
the last two blocks are merged.
Assuming the input is already sorted by release time, \ALG\ runs in
$O(n)$ time since each job ceases to be the first job of a block once.

\subsection{Finding all non-dominated schedules}

\begin{figure}
\begin{center}
\epsfig{file=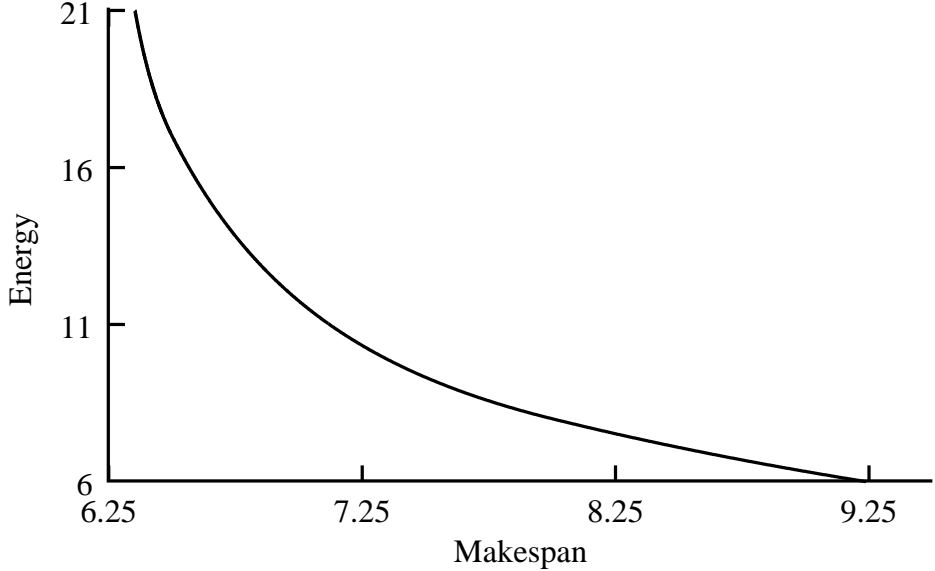,height=3in}
\end{center}
\vspace*{-2em}
\caption{Relationship
  between energy and makespan in non-dominated schedules for instance
  with $r_1=0$, $\work_1=5$, $r_2=5$, $\work_2=2$, $r_3=6$,
  $\work_3=1$, and $\text{power}=\text{speed}^3$.}
\label{makespan-energy-fig}
\end{figure}

\begin{figure}
\begin{center}
\epsfig{file=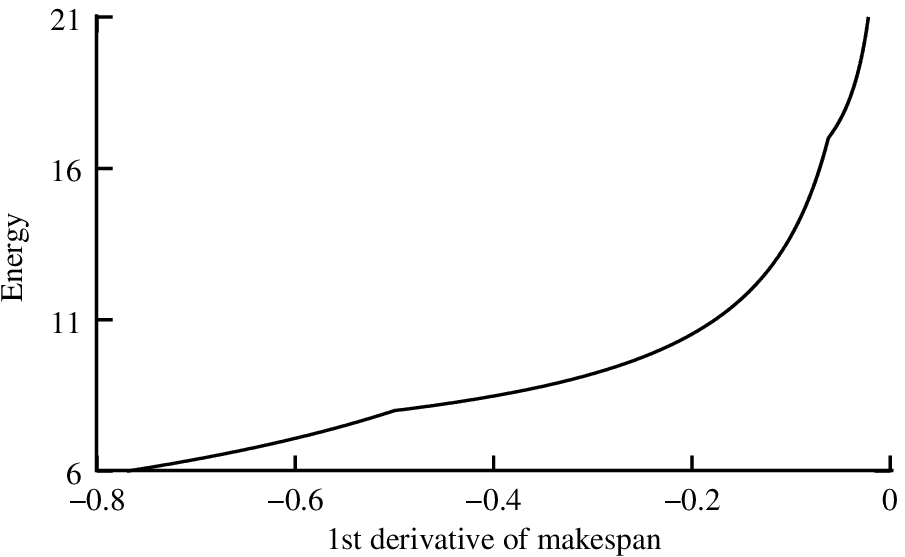,height=3in}
\end{center}
\vspace*{-2em}
\caption{Relationship between energy and 1st derivative of makespan
  in non-dominated schedules for instance with $r_1=0$, $\work_1=5$,
  $r_2=5$, $\work_2=2$, $r_3=6$, $\work_3=1$, and
  $\text{power}=\text{speed}^3$.}
\label{dmakespan-energy-fig}
\end{figure}

\begin{figure}
\begin{center}
\epsfig{file=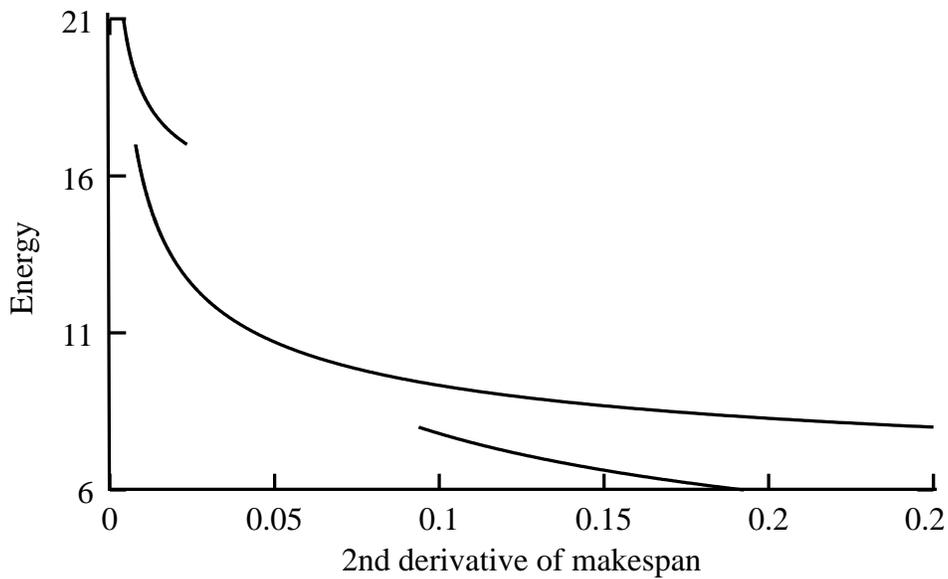,height=3in}
\end{center}
\vspace*{-2em}
\caption{Relationship between energy and 2nd derivative of makespan
  in non-dominated schedules for instance with $r_1=0$, $\work_1=5$,
  $r_2=5$, $\work_2=2$, $r_3=6$, $\work_3=1$, and
  $\text{power}=\text{speed}^3$.}
\label{ddmakespan-energy-fig}
\end{figure}

A slight modification of \ALG\ finds all non-dominated schedules.
Intuitively, the modified algorithm enumerates all optimal
configurations (i.e. ways to break the jobs into blocks) by starting
with an ``infinite'' energy budget and gradually lowering it.
To start this process, run \ALG\ as above, but omit the
merging step for the last job, essentially assuming the energy
budget is large enough that the last job runs faster than its predecessor.
To find each subsequent configuration change, calculate the energy budget at
which the last two blocks merge.
Until this value, only the last block changes speed.
Thus, we can easily find the relationship between makespan and energy
consumption for a single configuration and the curve of all
non-dominated schedules is constructed by combining these.
The curve for an instance with three jobs and
$\text{power}=\text{speed}^3$ is plotted in
Figure~\ref{makespan-energy-fig}.
The configuration changes occur at energy 8 and 17, but they are not
readily identifiable from the figure because the makespan/energy curve
is always continuous and has a continuous first derivative for this
power function.
Higher derivatives are discontinuous at the configuration changes.
Figures~\ref{dmakespan-energy-fig} and \ref{ddmakespan-energy-fig}
show the first and second derivatives.

\section{Impossibility of exactly minimizing flow \sloppy}
\label{impossibility-section}

We have completely solved uniprocessor power-aware makespan by 
showing how to compute all non-dominated schedules, forming a curve
such as Figure~\ref{makespan-energy-fig}.
The previous work on power-aware scheduling for
total flow~\cite{pruhs04b} includes a similar figure, but that figure
omits parts of the curve where the optimal schedule finishes one job
exactly as another is released.
We now show that these gaps cannot be filled exactly.

\begin{theorem} \label{impossible-theorem}
If $\text{power}=\text{speed}^3$, there is no exact algorithm to
minimize total flow for a given energy 
budget using operations
$+$, $-$, $\times$, $/$, and the extraction of roots, even on a
uniprocessor with equal-work jobs.
\end{theorem}

\begin{proof}
We show that a particular instance cannot be solved exactly.
Let jobs $J_1$ and $J_2$ arrive at time 0 and job $J_3$ arrive at time
1, each requiring one unit of work.
We seek the minimum-flow schedule using 9 units of energy.
Again we use $\speed_i$ to denote the speed of job $J_i$.
Thus,
\begin{equation}
\speed_1^2 + \speed_2^2 + \speed_3^2 = 9. \label{irreduc1-eqn}
\end{equation}
For energy budgets between approximately 8.43 and approximately 11.54,
the optimal solution finishes job $J_2$ at time 1.
Therefore,
\begin{equation}
\frac{1}{\speed_1} + \frac{1}{\speed_2} = 1 \label{irreduc2-eqn}
\end{equation}
and Theorem~\ref{flow-relationship-theorem} gives us that
\begin{equation}
\speed_1^3 = \speed_2^3 + \speed_3^3. \label{irreduc3-eqn}
\end{equation}
Substituting Equation (\ref{irreduc2-eqn}) into Equations
(\ref{irreduc1-eqn}) and (\ref{irreduc3-eqn}), followed by algebraic
manipulation gives
\begin{eqnarray*}
2\speed_2^{12}-12\speed_2^{11}+6\speed_2^{10}+108\speed_2^9-159\speed_2^8-738\speed_2^7+2415\speed_2^6
\makebox[1em]{}\\
-1026\speed_2^5-5940\speed_2^4
+12150\speed_2^3-10449\speed_2^2+4374\speed_2-729
& = & 0.
\end{eqnarray*}
According to the GAP system~\cite{gap}, the Galois group of
this polynomial is not solvable.
This implies the theorem by a standard result in
Galois theory (cf.~\cite[pg. 542]{dummit91}).
We owe the idea for this type of argument to Bajaj~\cite{bajaj88}.
\end{proof}
\vspace*{1em}

Since an arbitrarily-good approximation algorithm is known for total flow,
one interpretation of Theorem~\ref{impossible-theorem} is that
exact solutions do not have a nice representation.
For most applications, the approximation is sufficient since 
finite precision is the normal state of affairs in computer science.
Certainly, it could be used to draw an approximate curve for the gaps
in the flow analog of Figure~\ref{makespan-energy-fig}.
Only an exact algorithm such as \ALG\ can give closed-form solutions
suitable for symbolic computation, however.

\section{Multiprocessor scheduling} \label{multi-section}

Now we consider multiprocessor power-aware scheduling.
In a non-dominated schedule, the processors are
related by the following observations:
\begin{enumerate}
\item
For makespan, each processor must finish its last job at the same time
or slowing the processors that finish early would save
energy.
\item
For total flow, each processor's last job runs at the same speed or
running them at the average speed would save energy.
\end{enumerate}
Using these observations, slight modifications of \ALG\ and the total
flow algorithm of Pruhs et al.~\cite{pruhs04b} can solve
multiprocessor problems once the assignment of jobs to processors is
known.

We show how to assign equal-work jobs to processors for scheduling
metrics with two properties.
A metric is {\em symmetric} if it is not changed by permuting the
job completion times.
A metric is {\em non-decreasing} if it does not decrease when any
job's completion time increases.
Both makespan and total flow have these properties, but some metrics
do not.
One example is total weighted flow, which is not symmetric.

To prove our results, we need some notation.
For schedule $A$ and job $J_i$, let $\proc^A(i)$ denote the index of
the processor running job $J_i$ and $\next^A(i)$ denote
the index of the job run after $J_i$ on processor $\proc^A(i)$.
Also, let $\after^A(i)$ denote the portion of the schedule
running on processor $\proc^A(i)$ after the completion
of job $J_i$, i.e. the jobs running after job $J_i$ together with their
start and completion times.
We omit the superscript when the schedule is clear from context.

We begin by observing that job start times and completion times
occur in the same order.

\begin{lemma} \label{start-completion-order-lemma}
If $OPT$ is an optimal schedule for equal-work jobs under
a symmetric non-decreasing metric, then
$\start_i^{OPT} < \start_j^{OPT}$ implies
$C_i^{OPT} \leq C_j^{OPT}$.
\end{lemma}

\begin{proof}
Suppose to the contrary that $\start_i^{OPT} < \start_j^{OPT}$
and $C_i^{OPT} > C_j^{OPT}$.
Clearly, jobs $J_i$ and $J_j$ must run on different machines.
We create a new schedule $OPT'$ from $OPT$.
All jobs on machines other than $\proc(i)$ and $\proc(j)$
are scheduled exactly the same, as are those that run before jobs
$J_i$ and $J_j$.
We set the completion time of job $J_i$ in $OPT'$ to $C_j^{OPT}$ and the
completion time of job $J_j$ in $OPT'$ to $C_i^{OPT}$.
We also switch the suffixes of jobs following these two, i.e. run
$\after(i)$ on processor $\proc(j)$ and run
$\after(j)$ on processor $\proc(i)$.
Job $J_i$ still has positive processing time since
$\start_i^{OPT'}=\start_i^{OPT} < \start_j^{OPT} < C_j^{OPT} = C_i^{OPT'}$.
(The processing time of job $J_j$ increases so it is also
positive.)
Thus, $OPT'$ is a valid schedule.
The metric values for $OPT$ and $OPT'$ are the same since this
change only swaps the completion times of jobs $J_i$ and $J_j$.

We complete the proof by showing that $OPT'$ uses less energy than $OPT$.
Since the power function is strictly convex, it suffices to show that both
jobs have longer processing time in $OPT'$ than job $J_j$ did in $OPT$.
Job $J_j$ ends later so its processing time is clearly longer.
Job $J_i$ also has longer processing time since runs throughout the
time $OPT$ runs job $J_j$, but starts earlier.
\end{proof}
\vspace*{1em}

Using Lemma~\ref{start-completion-order-lemma}, we prove that an
optimal solution exists with the jobs 
distributed in {\em cyclic order}, i.e. job $J_i$ runs on processor
$(i \mod m)+1$.

\begin{theorem} \label{cyclic-theorem}
There is an optimal solution for equal-work jobs under any symmetric
non-decreasing metric with the jobs distributed in cyclic order.
\end{theorem}

\begin{proof}
Suppose to the contrary that no optimal schedule distributes the jobs
in cyclic order.
Let $i$ be the smallest value such that no optimal schedule
distributes jobs $J_1, J_2, \ldots, J_i$ in cyclic order and let $OPT$
be an optimal schedule that distributes the first $i-1$ jobs
in cyclic order.
To simplify notation, we create dummy jobs
$J_{-(m-1)}$, $J_{-(m-2)}$, \ldots, $J_0$, with job $J_{-(m-i)}$ assigned
to processor $i$.
By assumption, $\next(i-m) \neq i$.
Let $J_l$ be the job such that $\next(l)=i$, i.e. the job preceeding
job $J_i$.
Since the first $i-1$ jobs are distributed in cyclic order, if we
assume (without loss of generality) that jobs starting at the 
same time finish in order of increasing index,
then Lemma~\ref{start-completion-order-lemma} implies that
$C_{i-m}^{OPT} \leq C_l^{OPT}$.
(Details omitted.)

To complete the proof, we consider 3 cases.
In each, we use $OPT$ to create an optimal schedule assigning 
job $J_i$ to processor $(i \mod m)+1$,
contradicting the definition of $i$.

Case 1: Suppose no job follows job $J_{i-m}$.
We modify the schedule by moving $\after(l)$ to follow $J_{i-m}$ on
processor $(i \mod m)+1$.
Since $C_{i-m}^{OPT} \leq C_l^{OPT}$ and $\after(l)$ was able to follow
job $J_l$, it can also follow job $J_{i-m}$.
The resulting schedule has the same metric value and uses the same
energy so it is also optimal.

Case 2: Suppose $J_{i-m}$ is not the last job assigned to processor
$\proc(i-m)$ and $C_l^{OPT} < r_{\next(i-m)}$.
We extend the cyclic order by swapping $\after(l)$ and $\after(i-m)$.
This does not change the amount of energy used.
To show that it gives a valid schedule, we need to show that
jobs $J_l$ and $J_{i-m}$ complete before $\after(i-m)$ and $\after(l)$.
Job $J_l$ ends by time $\start_{\next(i-m)}^{OPT}$ by the assumption that
$C_l^{OPT} < r_{\next(i-m)}$.
Job $J_{i-m}$ ends by time $\start_{\next(l)}^{OPT}$ since
$C_{i-m}^{OPT} \leq C_l^{OPT}$.

Case 3: Suppose $J_{i-m}$ is not the last job assigned to processor
$\proc(i-m)$ and $C_l^{OPT} \geq r_{\next(i-m)}$.
In this case, we swap the jobs $J_{\next(i-m)}$ and $J_{\next(l)}=J_i$, but leave
the schedules the same.
In other words, we run job $J_{\next(i-m)}$ from time
$\start_{\next(l)}^{OPT}$ to time $C_{\next(l)}^{OPT}$ on processor
$\proc(l)$ and we run job $J_{\next(l)}$ from time 
$\start_{\next(i-m)}^{OPT}$ to time $C_{\next(i-m)}^{OPT}$ on processor
$\proc(i-m)$.
The schedules have the same metric value and each
uses the same amount of energy.
To show that we have created a valid schedule, we need to show that jobs
$J_{\next(i-m)}$ and $J_{\next(l)}$ are each released by the start
time of the other.
Job $J_{\next(i-m)}$ was released by time $\start_{\next(l)}^{OPT}$ since 
$C_l^{OPT} \geq r_{\next(i-m)}$.
Since a job with index greater than $i$ follows job $J_{i-m}$, 
$r_i = r_{\next(l)} \leq r_{\next(i-m)}$ and 
Job $J_{\next(l)}$ was released by time $\start_{\next(i-m)}^{OPT}$.
\end{proof}
\vspace*{1em}

A simpler proof suffices if we specify the makespan metric
since then $OPT$
has no idle time.
Thus, $r_{\next(i-m)} \leq C_{i-m}^{OPT} \leq C_l^{OPT}$ and
case 2 is eliminated.

Theorem~\ref{cyclic-theorem} allows us to solve multiprocessor
makespan for equal-work jobs.
Unfortunately, the general problem is NP-hard.

\begin{theorem}
Nonpreemptive
power-aware multiprocessor makespan is NP-hard, even when all jobs
arrive immediately.
\end{theorem}

\begin{proof}
We give a reduction from {\sc Partition}~\cite{garey79}:

\begin{quotation}
\noindent
{\sc Partition}:
Given a multiset $A = \{a_1, a_2, \ldots, a_n\}$,
does there exist a partition of $A$ into $A_1$ and
$A_2$ such that $\sum_{a_i \in A_1} a_i = \sum_{a_i \in A_2} a_i$?
\end{quotation}

Let $B=\sum_{i=1}^n a_i$.
We assume $B$ is even since otherwise no partition exists.
We create a scheduling problem from an instance of {\sc Partition} by
creating a job $J_i$ for each $a_i$ with $r_i=0$ and $\work_i=a_i$.
Then we ask whether a 2-processor schedule exists with
makespan $B/2$ and a power budget allowing work $B$ to run at speed 1.

From a partition, we can create a schedule where each processor runs the jobs
corresponding to one of the $A_i$ at speed 1.
For the other direction, the convexity of the power function implies
that all jobs run at speed 1 so the work must be partitioned between
the processors.
\end{proof}
\vspace*{1em}

Pruhs et al.~\cite{pruhs05} observed that the special case of all jobs
arriving immediately has a PTAS based on load balancing work by
Alon et al.~\cite{alon97} on minimizing the $L_\alpha$ norm of
loads.

\section{Future work} \label{conc-section}

The study of power-aware scheduling algorithms is just beginning so
there are many possible directions for future work.
We consider the most important to be finding online algorithms with
performance guarantees for makespan or total flow.
No such algorithms are currently known, but many scheduling
applications occur in the online setting.
Our results on the structure of optimal
solutions may help with this task, but the problem seems quite difficult.
If the algorithm cannot know when the last job has arrived, it
must balance the need to run quickly to minimize makespan if no other
jobs arrive against the need to conserve energy in case more jobs do
arrive.

We would also like to see theoretical research using models that more
closely resemble real systems.
With this objective, we have been investigating actual
implementations of dynamic voltage scaling.
The most obvious feature of real systems differing from the standard
model is that the speed has discrete settings rather than being a
continuous variable.
Imposing minimum and/or maximum speeds is one way to partially
incorporate this aspect of real systems without going all the way to
the discrete case.
Another feature of real systems is that slowing down the processor has
less effect on memory-bound sections of code since part of the running
time is caused by memory latency.
There is already some simulation-based work attempting to exploit
this phenomenon~\cite{xie03}.
Finally, real systems incur overhead to switch speeds because the processor
must stop while the voltage is changing.
This overhead is fairly small, but discourages algorithms requiring
frequent speed changes.
We have begun considering models incorporating some of these changes in
the hope of finding one that more closely reflects real systems
while remaining mathematically tractable.

\paragraph*{Acknowledgements}
We thank Jeff Erickson for introducing us to the work of
Bajaj~\cite{bajaj88} on using Galois theory to prove hardness results.
We also thank the anonymous referees for
pointing out an error in our discussion of related work and
acknowledge various helpful comments from Dan
Cranston, Erin Chambers, and Sariel Har-Peled.

\bibliographystyle{plain}

\end{document}